%Paper: 9204220
%From: "R. Sekhar Chivukula" <sekhar@weyl.bu.edu>
%Date: Mon, 13 Apr 92 15:21:24 -0400

%%%%%%%%%%%%Uses harvmac.tex %%%%%%%%%%%%%%%%%%%%%%%%%%%%%%%%%%%%%%%%%%%%%%%%
\input harvmac.tex

\noblackbox
\def\ie{{\it i.e.}}
\def\eg{{\it e.g.}}

\def\tr{\hbox{ tr }}

\def\half{{1 \over 2}}
\def\halfi{{i \over 2}}

\def\vevphij{\langle \phi_j \rangle}
\def\lint{\int {d^4 \ell \over (2 \pi)^4}}

\def\derLR{\rlap{\vbox to1em{%
     \hbox{$\scriptscriptstyle\leftrightarrow$}\vfil}}\partial}
\def\Boxmark#1#2#3{\global\setbox0=\hbox{\lower#1em \vbox{\hrule height#2em
     \hbox{\vrule width#2em height#3em \kern#3em \vrule width#2em}%
     \hrule height#2em}}%
     \dimen0=#2em \advance\dimen0 by#2em \advance\dimen0 by#3em
     \wd0=\dimen0 \ht0=\dimen0 \dp0=0pt
     \mkern1.5mu \box0 \mkern1.5mu }

\Title{\vbox{\baselineskip12pt\hbox{BUHEP-91-27}
  \hbox{hep-ph@xxx/9204220}}}
{Colored Pseudo-Goldstone Bosons and Gauge Boson Pairs}

\centerline{R. Sekhar Chivukula}
\vskip .1in
\centerline{Mitchell Golden}
\vskip .1in
\centerline{\it and}
\vskip .1in
\centerline{M. V. Ramana}
\vskip .1in
\centerline{\it Boston University}
\centerline{\it Dept. of Physics}
\centerline{\it 590 Commonwealth Avenue}
\centerline{\it Boston, MA 02215}
\vskip .2in
\centerline{\bf ABSTRACT}

If the electroweak symmetry breaking sector contains colored particles
weighing a few hundred GeV, then they will be copiously produced at a
hadron supercollider.  Colored technipions can rescatter into pairs of gauge
bosons.  As proposed by Bagger, Dawson, and Valencia, this leads to gauge
boson pair rates far larger than in the standard model.  In this note we
reconsider this mechanism, and illustrate it in a model in which the rates can
be reliably calculated.  The observation of both an enhanced rate of
gauge-boson-pair events and colored particles would be a signal that the
colored particles were pseudo-Goldstone bosons of symmetry breaking.

\Date{1/92}

\nref\techni{S. Weinberg, Phys. Rev. D {\bf 19}, 1277 (1979)\semi
L. Susskind, Phys. Rev. D {\bf 20}, 2619 (1979).}
\nref\DS{S. Dimopoulos and L. Susskind, Nucl. Phys. B {\bf 155}, 237 (1979).}
\nref\BDV{J. Bagger, S. Dawson, and G. Valencia, Phys. Rev. Lett. {\bf 67},
2256 (1991).}
\nref\CGS{R. S. Chivukula, M. Golden, and E. H. Simmons, Phys. Lett B {\bf
257}, 403 (1991); Nucl. Phys. B {\bf 363}, 83 (1991).}
\nref\CCWZ{S. Weinberg, Phys. Rev. {\bf 166}, 1568 (1968)\semi S.
Coleman, J. Wess, and B. Zumino, Phys. Rev. {\bf 177}, 2239 (1969).}
\nref\equival{J. Cornwall, D. Levin, and G. Tiktopoulos, Phys. Rev. D {\bf 10},
1145 (1974) \semi C. Vayonakis, Lett. Nuovo Cimento {\bf 17}, 383 (1976) \semi
M.~Chanowitz and M.~K.~Gaillard, Nucl. Phys. {\bf B261}, 379 (1985).}
\nref\SS{M. Soldate and R. Sundrum, Nucl. Phys. {\bf B340}, 1 (1990).}
\nref\CS{R. N. Cahn and M. Suzuki, Phys. Rev. Lett. 67, 169 (1991).}
\nref\Coleman{Sidney Coleman, R. Jackiw, and H. D. Politzer, Phys. Rev. {\bf
D10}, 2491 (1974).}
\nref\us{R. S. Chivukula and M. Golden, ``Scalar Resonances in Goldstone Boson
Scattering'', Boston University preprint BUHEP-91-11 \semi
R. S.  Chivukula and M. Golden, Phys. Lett. B {\bf 267}, 233 (1991).}
\nref\Einhorn{Martin B.  Einhorn, Nucl. Phys. B {\bf 246}, 75 (1984)\semi R.
Casalbuoni, D. Dominici, and R. Gatto, Phys. Lett. {\bf 147B}, 419 (1984).}
%\nref\pig{G. 't Hooft and M. Veltman, Nucl. Phys. B {\bf 153}, 365 (1991).}
\nref\virtstate{See, \eg, John R. Taylor, {\it Scattering Theory: The Quantum
Theory of Nonrelativistic Collisions}, John Wiley, New York, (1972)}
\nref\BroMik{R. Brown and K. Mikaelian, Phys. Rev. D {\bf 19}, 922 (1979).}
\nref\EHLQ{\message{I hate citing EHLQ!}E. Eichten, I. Hinchliffe, K. Kane,
and C. Quigg, Rev. Mod. Phys. {\bf 56}, 1 (1984).}

%\nfig\figone{}
\nfig\figtwo{The Feynman diagrams for $gg \to \phi\phi$ in the $O(N)$ model.}

\nfig\figthree{The $ZZ$ differential cross section in $nb/GeV$ vs $M_{ZZ}$
computed in the $O(N)$ model.  The solid and dashed lines show the $gg$ fusion
signal at the SSC and LHC respectively.  The $q\bar{q}$ background \BroMik\
are the dotdashed and dotted lines.  We have put $M=1800$GeV and $m_\psi =
120$GeV.  A rapidity cut of $|y|<2.5$ is imposed on the final state $Z$s. EHLQ
Set II structure functions \EHLQ\ are used.}

\vfill
\eject

Technicolor \techni\ remains an intriguing possiblility for electroweak
symmetry breaking.  Typically, the technihadrons which are lowest in mass are
the technipions.  A technicolor model must have at least three technipions,
because they become the longitudinal components of the $W$ and $Z$.  In
general there may be others.  In models where there are colored
technifermions, there are colored technipions.  The one-family model \DS, for
example, has 63 technipions, including some which tranform as $(8,3)$ under
$SU(3)_{color} \times SU(2)_{weak}$.

Colored technipions are easy to produce at a hadron supercollider.
Interestingly, in technicolor models a pair of colored technipions can
rescatter into the colorless ones.  For example, in the one-family model the
reaction $(8,3) (8,3) \to (1,3) (1,3)$ can occur.  Since the former particles
are easy to produce, and the latter are the longitudinal components of the
gauge bosons, there will be a large rate for gauge boson pair events at the SSC
or LHC.  This mechanism was recently proposed by Bagger, Dawson, and Valencia
\BDV\ to test electroweak symmetry breaking at a hadron supercollider.

We are left with an interesting possiblility.  The colored technipions are
produced in large numbers at the SSC and LHC, and should be observable even in
the worst case scenario in which they decay exclusively into light quarks and
flavor tagging is useless \CGS.  However, because they are produced strongly,
there is no obvious way to connect them to the symmetry breaking sector, and
indeed a skeptic might argue that the colored scalars are unrelated to
it\foot{On the other hand, the generic expectation is that the technipions
will mostly decay to the heaviest fermions.  Observation that the colored
scalar decayed in this way would be an argument that it had something to do
with flavor physics, and hence symmetry breaking.}.  It is the {\it
combination} of their discovery with the observation of a large number of
gauge-boson-pair events which permits us to argue that the colored scalars are
pseudo-Goldstone bosons of the symmetry breaking sector.

Since they are approximate Goldstone bosons, the low-energy behavior of the
technipions can be described by chiral Lagrangian techniques \CCWZ.  Consider
the lowest order, two-derivative, chiral Lagrangian for the one-family model:
$\CL_2 = (f^2/4) \tr [(D^\mu \Sigma)^\dagger (D_\mu \Sigma)]$, where $\Sigma =
\exp(2i \pi^a T^a / f)$.  Here $T^a$ are generators of $SU(8)$, and $\pi^a$
are the 63 technipion fields.  The derivatives above are gauge covariant,
using the proper imbedding of the color and electroweak groups into the chiral
$SU(8)_L \times SU(8)_R$.  We may fix $f$ by noting that since there are four
electroweak doublets condensing in this model -- three quarks and one lepton
-- we have $M_W^2 = g^2 f^2$.  From this we deduce that $f = v/2$, where $v
\approx 250$GeV.

This lowest order chiral Lagrangian contains terms which allow the computation
of the $g g \to Z Z$ process in which we are interested.  The computation is
facilitated by the use of the equivalence theorem \equival, which states that
at energies large compared to the $W$ mass, the amplitude for a process
containing external longitudinal gauge bosons is equal to the amplitude for the
process with the longitudinal gauge bosons replaced by their swallowed
unphysical Goldstone bosons.  It is impossible to construct a gauge and chiral
invariant four-derivative counterterm for the coupling of the gluons to the
uncolored swallowed technipions, so the one-loop calculation of the gluon-gluon
to the longitudinal $ZZ$ state must be finite.  This is the computation
presented by Bagger, Dawson, and Valencia \BDV.

To how high an energy scale can we trust the calculations using $\CL_2$?
Quite general considerations \SS\ show that in a theory in which the symmetry
breaking pattern is $SU(N)_L \times SU(N)_R \to SU(N)_V$, the scale at which
the calculations fail must fall as $1/\sqrt{N}$.  Consider $\pi^a \pi^b \to
\pi^c \pi^d$ scattering.  We can calculate the $SU(N)_V$-singlet $s$-wave
amplitude using $\CL_2$ \CS:
\eqn\asing{
a_{00} = {N s \over 32 \pi f^2}~,
}
where $s$ is the usual Mandelstam variable.  This amplitude grows bigger than
1 in magnitude at a scale $4 \sqrt{2 \pi} f / \sqrt{N}$.  In the one-family
model, this is a mere 440 GeV!  Thus, the lowest order chiral Lagrangian must
fail at a very low energy, at a scale which would be easily reachable by the
SSC or LHC.

It is now easy to see what goes wrong when we calculate $gg \to ZZ$ using
$\CL_2$.  Consider the computation of the imaginary part of the amplitude
\foot{ The amplitude given in \BDV\ has no imaginary part because that paper
considered the case of massless colored technipions. Colored technipions are
expected to have a mass in the hundreds of GeV.}.  Only when the two gluons
produce a pair of on-shell technipions can the diagram be cut in such a way as
to give a physical intermediate state, and therefore the imaginary part of the
amplitude comes from the production of on-shell colored technipions which
rescatter into the swallowed Goldstone bosons.  However, as we have noted,
this rescattering will violate unitarity.  The $SU(8)_V$-singlet spin-0 part
fails first, at 440GeV.  We therefore conclude that the lowest-order chiral
Lagrangian calculation of $gg \to ZZ$ displays bad high-energy behavior.

The failure of the chiral Lagrangian to address the scattering of technipions
in the one-family model at the requisite energies leads us to consider a
simpler -- but calculable -- model which resembles technicolor.  Our toy model
of the electroweak symmetry breaking sector is based on an $O(N)$ linear
sigma-model solved in the limit of large $N$ \Coleman.  It has both exact
Goldstone bosons (which will represent the longitudinal components of the $W$
and $Z$) and colored pseudo-Goldstone bosons.  To this end let $N=j+n$ and
consider the Lagrangian density \us
\eqn\lnought{
{\cal L} = \half (\partial \vec{\phi})^2 +
\half (D \vec{\psi})^2 - \half \mu_{0\phi}^2 \vec{\phi}^2 - \half
\mu_{0\psi}^2 \vec{\psi}^2 - {\lambda_0 \over 8 N} {(\vec{\phi}^2 +
\vec{\psi}^2)}^2 ~,
}
where $\vec{\phi}$ and $\vec{\psi}$ are $j$- and $n$-component real vector
fields. This theory has an approximate $O(j+n)$ symmetry which is broken to
$O(j) \times O(n)$ so long as $\mu_{0\phi}^2 \neq \mu_{0\psi}^2$.  If
$\mu_{0\phi}^2$ is negative and less than $\mu_{0\psi}^2$, one of the
components of $\vec{\phi}$ gets a vacuum expectation value (VEV), breaking the
approximate $O(N)$ symmetry to $O(N-1)$.  With this choice of parameters, the
exact $O(j)$ symmetry is broken to $O(j-1)$ and the theory has $j-1$ massless
Goldstone bosons and, at tree level, one massive Higgs boson.  The $O(n)$
symmetry is unbroken, and there are $n$ degenerate massive pseudo-Goldstone
bosons.  We will consider this model in the limit that $j,n \rightarrow
\infty$ with $j/n$ held fixed.

The scalar sector of the standard one-doublet Higgs model has a global $O(4)
\approx SU(2) \times SU(2)$ symmetry, where the 4 of $O(4)$ transforms as one
complex scalar doublet of the $SU(2)_W\times U(1)_Y$ electroweak gauge
interactions. We will model the $O(4)$ of the standard model by the $O(j)$ of
the $O(j+n)$ model solved in the large $j$ and $n$ limit.  Of course, $j=4$ is
not particularly large.  Nonetheless, the resulting model will have the same
qualitative features, and we can investigate the theory at moderate to strong
coupling \Einhorn.

We have gauged an $SU(3)_c$ subgroup of $O(n)$, so the $\psi$ fields are
colored\foot{Technically, this gauging of the color symmetry for $\psi$ and
not $\phi$ breaks the $O(N)$ by a dimension-four operator.  We are free to
ignore this if we regard $\alpha_s$ as small compared to $\lambda$ and
accordingly neglect diagrams with loops involving gluons.}.  We have chosen
$\psi$ to be three color octets, analogous to the $(8,3)$ of the one-family
model.  Our choice corresponds to $n=24$.

A simple trick \Coleman\ for the solution of this theory to leading
order in $1/N$ involves introducing a new field $\chi$, and modifying the
Lagrangian to
\eqn\newL
{
\CL \to \CL + \half {N \over \lambda_0} {\left(\chi -
\half {\lambda_0 \over N} (\vec{\phi}^2 + \vec{\psi}^2)
- \mu_{0\phi}^2 \right)}^2~.
}
Adding this term has no effect on the dynamics of the theory: since the
added term has no space-time derivative, the path integration over the field
$\chi -
\half {\lambda_0 \over N} (\vec{\phi}^2 + \vec{\psi}^2 ) - \mu_{0\psi}^2$ will
yield an irrelevant overall constant.  On the other hand, the Feynman rules
generated from the new Lagrangian are different, since
\eqn\newLf{ {\cal L} =
\half (\partial\vec{\phi})^2 + \half (D \vec{\psi})^2
- { m_\psi^2 \over 2} \vec{\psi}^2
+ \half {N \over \lambda_0} \chi^2
- \half \chi (\vec{\phi}^2+ \vec{\psi}^2)
- {N \mu_{0\phi}^2 \over \lambda_0} \chi
}
where $m_\psi^2=\mu_{0\psi}^2-\mu_{0\phi}^2$ is a positive number and where we
have ignored an irrelevant constant.  Written in this way, the only nontrivial
interactions are the $\chi \phi^2$ and $\chi \psi^2$ terms.  The advantage of
this formalism is that when a diagram is evaluated, the only source of factors
of $1/N$ will be the $\chi$ propagators.  To evaluate any process to leading
order in $1/N$, therefore, one computes only diagrams with the minimum number
of $\chi$ propagators, \ie\ diagrams with no $\chi$ loops.

The gauge boson production diagrams of this model are shown in \figtwo.  The
double line indicates the $\chi$ propagator, $D_{\chi\chi}(s)$, including all
radiative corrections coming from loops of $\phi$s and $\psi$s.  Postponing
for the moment the evaluation of $D_{\chi\chi}(s)$, we see that this process
is very easy to calculate, since the $g\psi\psi$ vertex is just the ordinary
coupling of a gauge boson to a scalar, and the couplings of the $\chi$ to
$\phi\phi$ and $\psi\psi$ are both just $-i$.  We find that the sum of the
diagrams in \figtwo\ is
\eqn\triang{
{i \over 16 \pi^2} \left(g^{\mu\nu} - {2 p_2^\mu p_1^\nu \over s}\right)
2 n_8 C_8 I(s, m_\psi^2) D_{\chi\chi}(s)~.
}
Here $p_1$ and $p_2$ are the momenta of the two incoming gluons; their
polarization vectors are associated with the indices $\mu$ and $\nu$
respectively.  The number of octets in $\psi$ is $n_8$; as we noted above, we
have chosen $n_8 = 3$.  The factor $C_8$ denotes the Casimir operator of an
$SU(3)_c$ octet, which is 3.  The variable $s$ is $2 p_1 \cdot p_2$, and the
function $I(s, m_\psi^2)$ is a Feynman parameter integral
\eqn\Idef{
I(s, m_\psi^2) \equiv
\int_0^1 d\!x \int _0^{1-x} d\!y {xys \over m_\psi^2 - xys - i \epsilon}~.
}
%We may write this integral in terms of Spence functions using some simple
%manipulations \pig.

At this point, all that remains is to evaluate $D_{\chi\chi}(s)$.  The details
of its calculation may be found in \us.  Only one subtlety need concern us
here. In the process of solving the theory \lnought\ to leading order in
$1/N$, there are divergences which must be regularized and $\lambda$ and $\mu$
must be renormalized.  This can be accomplished by defining
\eqn\lambdaR{
{1 \over \lambda(M)} \equiv
{1 \over \lambda_0} - \halfi \lint {1 \over (\ell^2 + i \epsilon)
(\ell^2 - M^2 + i \epsilon)}~,
}
and
\eqn\muR{
{\mu^2(M) \over \lambda(M)} \equiv
{\mu_{0\phi}^2 \over \lambda_0} + \halfi \lint {1 \over \ell^2 + i \epsilon}~,
}
where $M$ is an arbitrary renormalization point.  These two subtractions are
sufficient to render the theory finite to leading order in $1/N$.
To leading order in $1/N$, $m_\psi^2$ remains unrenormalized.

Specifying $\lambda(M)$ and $\mu^2(M)$ (as well as $m_\psi^2$) for a
particular $M$ specifies the theory completely.  We will choose $\mu^2(M)$
negative and $m_\psi^2>0$, so that the $O(j)$ symmetry is spontaneously broken
and we will orient the VEV of $\vec{\phi}$ so that only $\vevphij \neq 0$.
Instead of $\mu^2(M)$, we will work with the parameter $\vevphij$ directly,
since it has physical meaning and the coefficients in the Lagrangian do not.
Consider the $O(N)$ symmetry current $J^\nu_\alpha = i \vec{\phi}^T
{\derLR}^\nu T_\alpha \vec{\phi}/2$ where $T_\alpha$ is a generator of $O(N)$,
normalized to tr$T_\alpha T_\beta = 2 \delta_{\alpha \beta}$.  When $\phi_j$
gets a VEV, the broken symmetry currents will satisfy $J^\nu_a = i \vevphij
\partial^\nu \phi_a + \ldots$, and so we identify $\vevphij = v$.

At this point we may trade in the parameter $\lambda$ for the scale $M$.
Instead of regarding the renormalization point $M$ as fixed and $\lambda$ as
varying, we take
\eqn\transmut{
{1 \over \lambda(M)} = 0
}
and, therefore, $M$ specifies the strength of the coupling.  Of course
equation \transmut\ implies that $M$ is the scale at which the renormalized
coupling blows up.  The fact that the coupling becomes infinite at some finite
scale $M$ is a reflection of the triviality of the scalar $O(N)$ theory.
For the purposes of this work, however, this need not trouble us. The $O(N)$
model is a consistent effective theory for energies well below $M$ and we
will only need results in this regime.

To leading order in $1/N$, we find
\eqn\Dchichi{
D_{\chi\chi}(s) = {-is\over
{\left[ v^2 - Ns\left({1\over\lambda(M)}
+\widetilde{B}(s;m_\psi,M)\right)\right]}}~,
}
where
\def\xx{\sqrt{s / (4 m_\psi^2 - s)}}
\eqn\Btilde{
\eqalign{
\widetilde{B}(s; m_\psi, M)  = {n \over 32 N \pi^2}
& \left\{
 1
+ {i \over \xx} \log{i - \xx \over i + \xx}
- \log{m_\psi^2 \over M^2}
\right\}\cr
&+
{ j \over 32 N \pi^2 }
\left\{ 1 + \log{M^2 \over -s}
\right\}~.\cr
}
}

The logs and the square roots have branch cuts, and it is up to us to place
them in physically meaningful places.  This we do by considering $\phi \phi
\to \phi \phi$ scattering.  In \newL\ this proceeds entirely by entirely via
the $\chi$ exchange; we find that the $O(j-1)$ singlet spin zero scattering
amplitude is
\eqn\amp{
a(s) = {i j \over 32 \pi} D_{\chi\chi}(s)~.
}
This amplitude has two branch cuts just below the real axis \virtstate: one
starting at $s=0$ from pairs of massless Goldstone bosons, and the other
starting at $s=4m_\psi^2$ from pairs of pseudo-Goldstone bosons.  To leading
order in $1/N$, there are no other multiparticle states.  We have written
\Btilde\ so that these branch cuts are obtained using the conventional
definition of the log and the square root, in which the branch cut is just
under the negative real axis.  That is, $\log z = \log |z| + i \theta$ and
$\sqrt{z} = |z|^{1/2} \exp{(i\theta/2)}$ where $-\pi < \theta \le \pi$.

In \figthree\ we show the $ZZ$ cross sections at hadron supercolliders.  Our
choice of $M = 1800$ GeV, and $m_\psi = 120$ GeV gives a strongly coupled,
QCD-like scattering amplitude\foot{For larger values of $M$ the amplitude is
weakly coupled, and there is a narrow Higgs resonance.  While this is also a
plausible scenario, it is not necessarily what we expect in a technicolor
theory.  See \CS\ and \us\ for a discussion of this point.}.  Since we have
used the equivalence theorem, the $Z$ mass is ignored in the amplitude.
However, the $Z$ mass has been retained in the phase space.  In this
process the cross section for the $W^+W^-$ final state is double that of $ZZ$.

There are two interesting features of these graphs.  First, we note that the
cross sections fall rapidly at high energies.  In the computation using the
lowest order chiral Lagrangian the high energy behavior is quite different -
the figures in \BDV\ show that the differential cross sections fall by less
than a factor 4 between $M_{ZZ} = 200$ and 1000 GeV.  As we have noted, this
is because the technipion rescattering cannot be as large as given by $\CL_2$.
In contrast, in the toy model the Goldstone boson $S$ matrix is unitary at all
energies to leading order in $1/N$.  Accordingly, the amplitude (\triang\ and
\Dchichi) for $gg \to ZZ$ will not display the bad high-energy behavior
$\CL_2$.  In fact, \figthree\ shows that at high energies the background cross
section exceeds that of the signal.  This is not surprising, because the $gg$
luminosity falls considerably faster than that of $q\bar{q}$.

The second interesting feature of these graphs is that the cross sections are
quite large, and should easily be observable.  These cross sections are much
larger than those of non-resonant $ZZ$ production via a top-quark loop.
Accordingly, we neglect the top contribution to this process.  One should not
trust the numbers on these graphs too much -- the true calculation is model
dependent and one certainly should not take the toy model too seriously.
Nonetheless, it is plausible that the computation we have done is
conservative, since there may be more colored pseudo-Goldstone bosons than we
have assumed, or there may be some in representations with higher Casimirs.

We conclude by noting that the following interesting scenario may be observed
at a hadron supercollider.  There could be be colored, weakly-decaying
particles produced in great numbers.  However, there might be no obvious way
to connect them to symmetry breaking.  Though the process cannot be computed
using the chiral Lagrangian, there will be rescattering of these particles
into longitudinal Goldstone bosons.  If we observe colored particles in
conjunction with very large rates of electroweak-gauge-boson pair events, then
we may have a hint that they are pseudo-Goldstone bosons.

We would like to thank J.~Bagger, A.~Cohen, S.~Dawson, M.~Dugan, and G.
Valencia for useful conversations, and K.~Lane and S.~Selipsky for reading the
manuscript.  R.S.C.  acknowledges the support of an NSF Presidential Young
Investigator Award and of an Alfred P.  Sloan Foundation Fellowship.  M.G.
thanks the Texas National Research Laboratory Commission for support under a
Superconducting Super Collider National Fellowship.  This work was supported
in part under NSF contract PHY-9057173 and under DOE contract
DE-AC02-89ER40509 and by funds from the Texas National Research Laboratory
Commission under grant RGFY91B6.

\listrefs
\listfigs

\bye